\documentclass[11pt]{article}

\makeatletter 

\usepackage{color}
\usepackage{amsmath,amssymb}
\usepackage{wrapfig,url,slashed,cite}
\usepackage{hyperref}    
\usepackage{graphicx}

\renewcommand\citepunct{,\penalty\@M\hskip.13emplus.1emminus.1em\relax} 
\usepackage{enumerate}
\usepackage{feynmp}
\hypersetup{pdfauthor={Motoi Endo and Sho Iwamoto},pdftitle={LHC Dijet Signals in New Physics Models for Top Forward--Backward Asymmetry}}

\newcommand{\dummybibliography}[1]{\relax}

\newcommand{\dd}{{\mathrm d}}      

\newcommand\Order{\mathop{\mathcal{O}}}

\newcommand{\@diff}   [4]{\dfrac{#4#3#1}{#4#2#3}}
\newcommand{\diff}    [2]{\@diff{#1}{#2}{}{\dd}}
\newcommand{\pdiff}   [2]{\@diff{#1}{#2}{}\partial}
\newcommand{\fdiff}   [2]{\@diff{#1}{#2}{}{\delta}}
\newcommand{\ndiffnum}[3]{\@diff{#1}{#2}{^#3}\dd}
\newcommand{\npdiff}  [3]{\@diff{#1}{#2}{^#3}\partial}
\newcommand{\@difftwo}[4]{\dfrac{#4^2#1}{#4#2\,#4#3}}
\newcommand{\difftwo} [3]{\@difftwo{#1}{#2}{#3}{\dd}}
\newcommand{\pdifftwo}[3]{\@difftwo{#1}{#2}{#3}{\partial}}

\newcommand{\un}[1]{{\mathrm{\,#1}}} 
\newcommand{\TeV}{\un{TeV}}
\newcommand{\GeV}{\un{GeV}}
\newcommand{\MeV}{\un{MeV}}
\newcommand{\keV}{\un{keV}}

\def\@xxxEV{\@ifnextchar-{\@xxxEV@minus}{\@xxxEV@plus}}
\def\@xxxEV@plus#1#2{%
  \ifnum{#1=0}{}\else\ifnum{#1=1}{10}\else {10^#1}\fi\fi #2}
\def\@xxxEV@minus#1#2 {10^{-#1}{\rm\,#2}}

\newcommand{\TEV}[1]{\@xxxEV{#1}{\TeV}}
\newcommand{\GEV}[1]{\@xxxEV{#1}{\GeV}}
\newcommand{\MEV}[1]{\@xxxEV{#1}{\MeV}}
\newcommand{\KEV}[1]{\@xxxEV{#1}{\keV}}

\def\EE{\@ifnextchar-{\@@EE}{\@EE}}
\def\@EE#1{\ifnum#1=1\times10\else\times10^{#1}\fi}
\def\@@EE#1#2{\!\times\!10^{-#2}}

\def\T{\@ifnextchar^{\T@u}{\@ifnextchar_{\T@d}{}}}
\def\T@u^#1{{^{#1}}\T}
\def\T@d_#1{{_{#1}}\T}

\let\gsim\gtrsim
\let\ol\overline

\newcommand{\s}[1]{_\mathrm{#1}}    

     \newcommand{\bL}{b\s L}

\newcommand{\invfb}{\un{fb^{-1}}}
\newcommand{\invpb}{\un{pb^{-1}}}
\newcommand{\fb}{\un{fb}}
\newcommand{\ECM}{E\s{CM}}
\newcommand{\Lag}{\mathcal{L}}
\newcommand{\PR}{P\s R}

\newcommand{\ttbar}{{t\bar t}}
\newcommand{\ppbar}{{p\bar p}}
\newcommand{\afb}{A\s{FB}}

\newcommand{\sigmadijet}{\sigma_{pp\to {\rm jj}}}

\makeatother 

\begin{document}
\baselineskip=18pt

\begin{titlepage}

\begin{flushright}
UT--11--33\\
IPMU--11--0167
\end{flushright}

\vskip 1.35cm
\begin{center}
{\Large \bf
LHC Dijet Signals in New Physics Models \\ for Top Forward--Backward Asymmetry
}
\vskip 1.2cm
Motoi Endo$^{1,2}$, Sho Iwamoto$^{1}$
\vskip 0.4cm

{\it $^1$  Department of Physics, University of Tokyo,
Tokyo 113--0033, Japan\\
$^2$ Institute for the Physics and Mathematics of the Universe,
University of Tokyo,\\ Chiba 277--8568, Japan
}

\vskip 2cm

\abstract{
The dijet signature at the LHC is studied in new physics models for the top forward--backward asymmetry at the Tevatron. In the $t$-mediator models, flavor-changing interactions contribute to the dijet production cross section as well as the asymmetry at least at the one-loop level. It is found that the LHC dijet measurements at 36pb$^{-1}$ have constrained the $Z'$ coupling larger than 2.5--3. The sensitivity is expected to be improved significantly as the integrated luminosity increases in the LHC.
}
\end{center}
\end{titlepage}

\setcounter{footnote}{0}
\setcounter{page}{2}

\section{Introduction}

Anomalies of the forward--backward (FB) asymmetry of top quark pair production have been reported by 
the Tevatron experiments. The asymmetry is measured by reconstructing the $t\bar t$ events from 
the semi-leptonic decay channels, where either $t$ or $\bar t$ decays leptonically and the other hadronically, or the di-lepton channels. The rapidity 
distribution of the reconstructed top and anti-top quarks defines the FB asymmetry,
\begin{align}
 \afb = \frac{\text{\#events}(\Delta y>0) - \text{\#events}(\Delta y<0)}{\text{\#events}(\Delta y>0) + \text{\#events}(\Delta y<0)},
\end{align}
where $\Delta y$ is the rapidity difference between $t$ and $\bar t$. The measured data depend on 
details of detector. After correcting for (unfolding) detector effects and acceptance, the asymmetry is 
inferred at the parton (production) level, which is comparable to the Standard Model (SM) prediction. 
The measurement of the discrepancy of the experimental data from the SM value is a key to uncover
evidence for physics beyond the SM.

The FB asymmetry has been measured in several decay modes. The $t\bar t$ reconstruction based on
the semi-leptonic channels provides the inclusive parton-level FB asymmetry
$\afb = 0.158 \pm 0.072 \pm 0.017$ from the CDF experiment~\cite{Aaltonen2011emd},
and $\afb = 0.196 \pm 0.060 ^{+0.018}_{-0.026}$ from the D0~\cite{Abazov2011fai},
which correspond to the integrated luminosity 5.3$\invfb$ and $5.4\invfb$ respectively.
The CDF also measured the asymmetry in the di-lepton channel, $\afb = 0.42 \pm
0.15 \pm 0.05$, at the parton level with 5.1$\invfb$~\cite{CDF10436}. Averaging these three results, the asymmetry becomes
\begin{align}
  \afb = 0.20 \pm 0.05~~~~~({\rm CDF+D0}),
\label{eq:afb_combined}
\end{align}
where the statistical and systematic uncertainties are combined in quadrature. 
This is compared with the SM prediction, 
\begin{align}
  \afb = 0.0724^{+0.0106}_{-0.0072}~~~~~({\rm SM}), 
\end{align}
at the NLO+NNLL 
level with {\tt MSTW2008} set of the Parton Distribution Functions (PDFs) and $m_t = 173.1$ GeV~\cite{Ahrens2011tfa}. Thus, the experimental result is larger than the SM prediction at the 2.5$\sigma$ level. 
Although electroweak corrections enhance the SM value by $\sim 10$\%~\cite{Hollik2011ect}, the discrepancy 
is still larger than 2$\sigma$.\footnote{
The discrepancy may be enhanced in specific regions of $\Delta y$ and/or $M_\ttbar$. 
The experimental results tend to show a larger asymmetry for large $M_\ttbar$, though 
it agrees with the SM prediction if the CDF and D0 data are combined.}

A large asymmetry has been reported from the D0 in the leptonic channels of the $t\bar t$ decays. 
Defining the leptonic asymmetry $\afb^{\ell,\ppbar}$ in terms of the event numbers for $q_\ell y_\ell$, where 
$q_\ell$ and $y_\ell$ are the charge and rapidity of the lepton, the data show~\cite{Abazov2011fai}
\begin{align}
  \afb^{\ell,\ppbar} = 0.152 \pm 0.038 ^{+0.010}_{-0.013}~~~~~({\rm D0})
\end{align}
at the parton level. This is deviated by more than the 3$\sigma$ level from the SM prediction, $\afb^{\ell,\ppbar}
= 0.021 \pm 0.001$, which is estimated at the NLO level by using the MC@NLO package~\cite{MC@NLO}.

These large FB asymmetries may indicate contributions to the top sector from physics beyond the SM 
(BSM models). It is important to study such BSM models from the aspect of LHC signatures.
In this letter, we focus on measurements of the dijet production cross section at the LHC.

For scenarios where the large FB asymmetries are 
sufficed by mediating a new particle at the $s$-channel between $u\bar u$ (or $d\bar d$) and 
$t\bar t$ ($s$-channel mediator models~\cite{Djouadi2010fba,Frampton2010aap}), the dijet measurements have provided 
a constraint at the tree level, leading to predictions of characteristic structures such as couplings or 
decay width of the mediator (see e.g.~\cite{aad2011snp}). 

If the asymmetry is due to $t$-channel exchange of a new particle which couples to the top and 
light quarks ($t$-channel mediator models~\cite{Jung2010tqf,Cheung2009tqf,Frampton2010aap,Arhrib2010fba,Dorsner2010lcs,Barger2011tat}), the dijet measurement is not restrictive at the tree level as long as the mediator does not couple to a pair of the light quarks. However, the relevant flavor changing interactions generally contribute to the dijet cross section, $\sigma_{pp \to jj}$, in loop levels (see diagrams in Fig.~\ref{fig:dijetdiagram}).
In this letter, we study signature of the $t$-channel mediator models in the LHC dijet cross section measurement at the one-loop level.

\section{Models}

The top FB asymmetry has been studied extensively in BSM models (see \cite{Aguilar-Saavedra2011smt,Gresham2011tta} for comprehensive studies, and \cite{Grinstein2011FSS} for models with flavor symmetry).
The $t$-channel (or $u$-channel) mediator models are phenomenologically classified into the following two types: the vector and scalar mediator models.
The vector mediator which couples to the $u$ and $t$ quarks is known to be (flavor-changing) $Z'$, while that couples to $d$ instead of $u$ is the $W'$ boson.
These models are favored because they yield positive effects on the asymmetry.
The scalar mediators are characterized by the color charge of the mediator and the light quark to which the mediator couples.
In this letter, we concentrate on the vector models, and the scalar ones will be mentioned later.

\begin{figure}[t]
\begin{center}
\begin{fmffile}{feyn/tch_vec}
\begin{fmfgraph*}(100,80)\fmfstraight
\fmfleft{q2,q1}\fmfright{t2,t1}\fmf{plain}{q1,v1,t1}\fmf{plain}{q2,v2,t2}\fmf{boson,lab=$Z'\:(W')$,l.s=left}{v1,v2}
\fmfv{l=$u\:(d)$,l.a=180}{q1}\fmfv{l=$\bar u\:(\bar d)$,l.a=180}{q2}\fmfv{l=$t$,l.a=0}{t1}\fmfv{l=$\bar t$,l.a=0}{t2}\fmfdot{v1,v2}
\end{fmfgraph*}
\end{fmffile}
\end{center}
\caption{The LO diagram which contributes to $\afb$ in models with a $t$-channel vector mediator. The initial quarks are $u$-quarks for $Z'$-model and $d$-quarks for $W'$-model.}
\label{fig:VectorMediatedTchannel}
\end{figure}
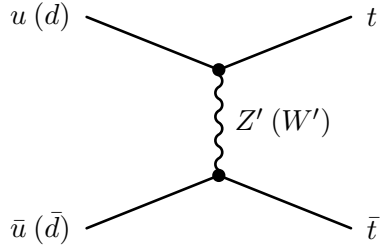

\subsection{Flavor changing $Z'$ model}
The flavor changing $Z'$ boson model is proposed as\cite{Jung2010tqf}
\begin{equation}
\Lag\s{NP} = \left(\lambda Z'_\mu\bar u\gamma^\mu \PR t + \text{h.c.}\right) + \epsilon_U \lambda Z'_\mu\bar u_i\gamma^\mu \PR u_i,
\end{equation}
where $\lambda$ is a coupling constant which is relevant for the top FB asymmetry, originally called $g_X$ in Ref.~\cite{Jung2010tqf}, and the $\epsilon_U$ term is a small flavor-diagonal interaction ($\epsilon_U\ll 1$) to avoid like-sign top quark events, i.e. $p\bar p\ \to Z'Z' \to (\bar ut)(\bar ut)$.

In this letter, we will adopt this model but with ignoring $\epsilon_U$ term:
\begin{equation}
\text{$Z'$-model\ :}\quad
\Lag\s{NP} = \lambda \left(Z'_\mu\bar u\gamma^\mu \PR t + \text{h.c.}\right).
\end{equation}
This interaction contributes to the top FB asymmetry. 
The lowest-order (LO) contribution is presented in Fig.~\ref{fig:VectorMediatedTchannel}. 
This model also induces the dijet scattering, $uu \to uu$, in the LHC at the one-loop level. 
The LO contribution is $\Order(\lambda^4)$ and/or $\Order(\lambda^2g\s s^2)$.
We will focus on the parameter space where $\lambda$ is as large as $\Order(1)$, and thereby $\Order(\lambda^4)$ contributions dominate, which is shown in Fig.~\ref{fig:dijetdiagram}.

The $\epsilon_U$ term also contributes to the dijet process at the tree level, which interferes with the $\lambda$ contribution. 
Since this depends on details of the $\epsilon_U$ term, we will discard the contribution for simplicity.
We also ignore the coupling between $Z'$ and left-handed quarks.
This is because it accompanies a $Z'$--$\ol d\s L$--$\bL$ interaction, which must be small to avoid superfluous contributions to $B_d-\ol{B_d}$ mixing.
It also gives extra contributions to the dijet cross section and tightens the dijet constraints severe.
So it is well ignored as long as $\lambda$ is large.

\subsection{Flavor changing $W'$ model}
\begin{figure}[t]\begin{center}
\begin{fmffile}{feyn/dij_vec}
\begin{fmfgraph*}(100,80)\fmfstraight
\fmfleft{x1,a2,x2,x8,a1,x3}\fmfright{x4,b2,x5,x7,b1,x6}
\fmf{plain}{a1,v1}\fmf{plain,lab=$t$,l.s=left,l.d=3pt,t=0.5}{v1,v2}\fmf{plain}{v2,b1}
\fmf{plain}{a2,v3}\fmf{plain,lab=$t$,l.s=left,l.d=3pt,t=0.5}{v3,v4}\fmf{plain}{v4,b2}
\fmfv{l=$q$,l.a=180}{a1}\fmfv{l=$q$,l.a=180}{a2}\fmfv{l=$q$,l.a=0}{b1}\fmfv{l=$q$,l.a=0}{b2}
\fmffreeze\fmfdot{v1,v2,v3,v4}\fmf{boson}{v1,v3}\fmf{boson}{v2,v4}
\end{fmfgraph*}\hspace{60pt}
\begin{fmfgraph*}(100,80)\fmfstraight
\fmfleft{x1,a2,x2,x8,a1,x3}\fmfright{x4,b2,x5,x7,b1,x6}
\fmf{plain}{a1,v1}\fmf{plain,lab=$t$,l.s=left,l.d=3pt,t=0.5}{v1,v2}\fmf{plain}{v2,b1}
\fmf{plain}{a2,v3}\fmf{plain,lab=$t$,l.s=left,l.d=3pt,t=0.5}{v3,v4}\fmf{plain}{v4,b2}
\fmfv{l=$q$,l.a=180}{a1}\fmfv{l=$q$,l.a=180}{a2}\fmfv{l=$q$,l.a=0}{b1}\fmfv{l=$q$,l.a=0}{b2}
\fmffreeze\fmfdot{v1,v2,v3,v4}\fmf{boson}{v1,v4}\fmf{boson}{v2,v3}
\end{fmfgraph*}

\begin{fmfgraph*}(100,80)\fmfstraight
\fmfleft{x1,a2,x2,x8,a1,x3}\fmfright{x4,b2,x5,x7,b1,x6}
\fmf{plain}{a1,v1}\fmf{plain,lab=$t$,l.s=left,l.d=3pt,t=0.5}{v1,v2}\fmf{phantom}{v2,b1}
\fmf{plain}{a2,v3}\fmf{plain,lab=$t$,l.s=left,l.d=3pt,t=0.5}{v3,v4}\fmf{phantom}{v4,b2}
\fmfv{l=$q$,l.a=180}{a1}\fmfv{l=$q$,l.a=180}{a2}\fmfv{l=$q$,l.a=0}{b1}\fmfv{l=$q$,l.a=0}{b2}
\fmffreeze\fmfdot{v1,v2,v3,v4}\fmf{boson}{v1,v3}\fmf{boson}{v2,v4}\fmf{plain}{v2,b2}\fmf{plain}{v4,b1}
\end{fmfgraph*}\hspace{60pt}
\begin{fmfgraph*}(100,80)\fmfstraight
\fmfleft{x1,a2,x2,x8,a1,x3}\fmfright{x4,b2,x5,x7,b1,x6}
\fmf{plain}{a1,v1}\fmf{plain,lab=$t$,l.s=left,l.d=3pt,t=0.5}{v1,v2}\fmf{phantom}{v2,b1}
\fmf{plain}{a2,v3}\fmf{plain,lab=$t$,l.s=left,l.d=3pt,t=0.5}{v3,v4}\fmf{phantom}{v4,b2}
\fmfv{l=$q$,l.a=180}{a1}\fmfv{l=$q$,l.a=180}{a2}\fmfv{l=$q$,l.a=0}{b1}\fmfv{l=$q$,l.a=0}{b2}
\fmffreeze\fmfdot{v1,v2,v3,v4}\fmf{boson}{v1,v4}\fmf{boson}{v2,v3}\fmf{plain}{v2,b2}\fmf{plain}{v4,b1}
\end{fmfgraph*}
\end{fmffile}\end{center}
\caption{New vector boson contributions to the dijet cross section in pp collision, where $q$ is the $u$-quark for $Z'$ model and the $d$-quark for $W'$.}
\label{fig:dijetdiagram}
\end{figure}
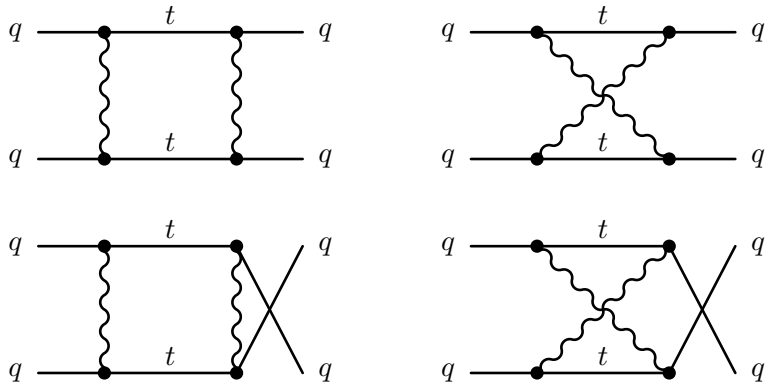

The Lagrangian of a flavor changing $W'$ boson model~\cite{Cheung2009tqf} is
\begin{align}
 \text{$W'$-model :}\quad&
 \Lag\s{NP} = \lambda \left(W'_\mu\bar d\gamma^\mu\PR t+\text{h.c.}\right).
\end{align}
where we ignore the interactions  between $W'$ and left-handed quarks as well as the $Z'$-model.
The LO contribution to $\afb$ in this model is similar to the $Z'$ case as shown in Fig.~\ref{fig:VectorMediatedTchannel}, and the diagrams of the dijet scattering are in Fig.~\ref{fig:dijetdiagram}.

\section{Observables}
We study the dijet process in the above two models.
These models involve flavor-changing interactions with the mediator, which are necessary to induce large $\afb$.
Under such interactions, the light quarks are scattered into a dijet final state by exchanging the mediator and the top quark at the radiative level, as shown in Fig.~\ref{fig:dijetdiagram}

The angular distribution of the inclusive dijet production cross section $\sigmadijet$ is sensitive to the 
new physics contributions.
Whereas the SM prediction of the dijet distribution for the angular variable $\chi \equiv (1+|\cos\theta|)/(1-|\cos\theta|)$ is relatively flat, the BSM contributions are likely to peak at low $\chi$. 
The CMS collaboration published two measurements for the dijet angular distribution of $pp$-collisions at $\ECM=7\TeV$~\cite{Khachatryan2011mod,Chatrchyan2011mod}, both of which correspond to $36\invpb$ experimental data, and both are in good agreement with the prediction of perturbative QCD.
The similar result was obtained by the ATLAS collaboration~\cite{aad2011snp}.
The present result and future expectation are discussed for the BSM models.

The inclusive FB asymmetry of the top quark is estimated in the same parameter space with assuming 
that the above (flavor-changing) Lagrangians dominate the new physics contributions to $\afb$.
Also, another constraint is considered in this letter: the differential cross section $\dd\sigma/\dd M_{t\bar t}$ of $t\bar t$ production in $p\bar p$ collisions, which is reported by the CDF collaboration~\cite{Aaltonen2009fmo} using integrated luminosity of $2.7\fb^{-1}$ at $\ECM=1.96\TeV$, and is consistent with the SM prediction.

\section{Method}
The BSM models are handled by utilizing {\tt FeynRules 1.5}~\cite{FeynRules}.
They are plugged to {\tt MadGraph 5.0}~\cite{MadGraph5} to generate events and calculate the asymmetry and the $\ttbar$ cross section.
The simulation is based on the LO QCD calculation, and {\tt CTEQ6L1}~\cite{PDFCTEQ6} is used as the PDF.

In the analysis, the $K$-factors are not introduced both for the SM and BSM cross sections.
Since the observables in this letter are defined by ratios of the cross sections except for $\dd\sigma/\dd M_{t\bar t}$, a part of NLO corrections are canceled between the SM and BSM.
On the other hand, since the NLO contributions would increase the $\ttbar$ cross section, the constraint 
from $\dd\sigma/\dd M_{t\bar t}$ is considered to be conservative.


In calculation of the dijet angular distribution we use {\tt FeynArts 3.6}~\cite{FeynArts}, {\tt FormCalc 7.1} and {\tt Looptools 2.6}~\cite{FormCalc} along with {\tt FeynRules}.
The SM contribution is calculated at the LO (tree) level with the PDF {\tt CTEQ6L1}, while the BSM contribution is calculated at one-loop level with the PDF {\tt CT10}~\cite{PDFCT10}.
The dijet angular distributions in the BSM models are tested by the $\chi^2$-method.
In the test, instead of comparison between the LHC experimental results and the BSM estimation, which needs full SM NLO calculation, we compare the BSM estimation (one-loop level BSM + tree level SM) with the tree-level SM calculated values for simplicity, using uncertainties borrowed from the experimental results.
To validate this procedure, it is checked that the SM LO calculation matches the experimental results.
Although there remain certain deviations between the SM LO and experimental values especially in small rapidity regions, which can be sufficed by NLO correction, the CMS experimental results \cite{Khachatryan2011mod,Chatrchyan2011mod} are reproduced by this method.

The present CMS results correspond only to $36\invpb$ data, while the integrated luminosity is growing rapidly.
Such a large set of data reduces uncertainties and improves the sensitivity to the BSM contributions.
Thus, we additionally derive an expectation of the dijet constraint for $5\invfb$ data by reducing the statistical uncertainties by the factor $\sqrt{5\invfb/36\invpb} \simeq 11.8$, whereas the systematic uncertainties are, conservatively, kept the same as the $36\invpb$ case.

\section{Results}
\subsubsection{$Z'$ model}
\begin{figure}[p]\begin{center}
\includegraphics[width=230pt]{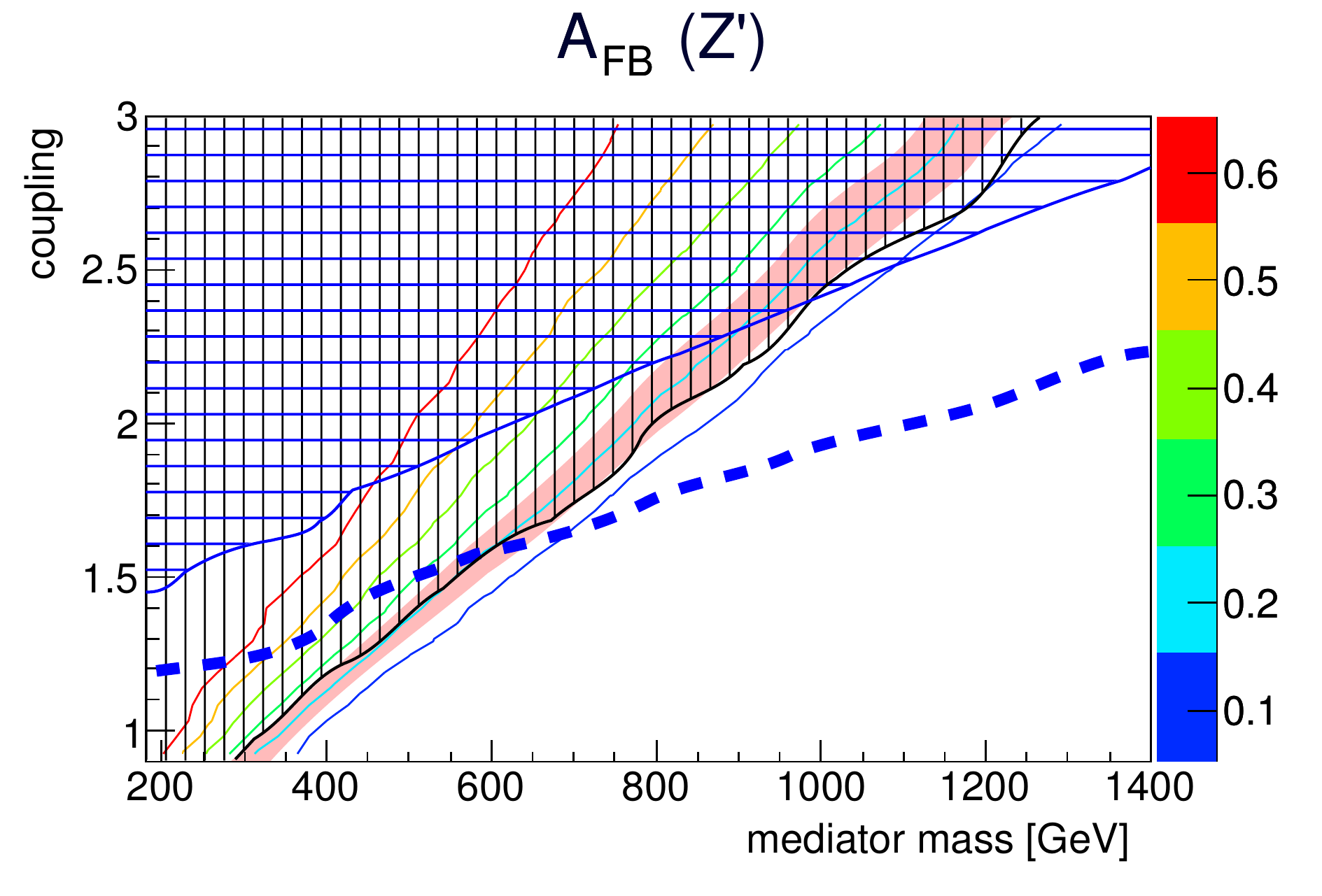}
 \caption{$\afb$ for the $Z'$-model. Experimental value Eq.~\eqref{eq:afb_combined} ($\pm1\sigma$) is the pink region. Rejected region from $\dd\sigma_{\ttbar}/\dd M_{\ttbar}$ (with no $K$-factor) is shown as black-stripe, and dijet constraints is as blue-border. Blue-dashed line shows an expected dijet constraint for a $5\invfb$ data of the CMS, where the region above the line will be covered.}
 \label{fig:exz_afb}
\end{center}\end{figure}
\begin{figure}[p]\begin{center}
\includegraphics[width=230pt]{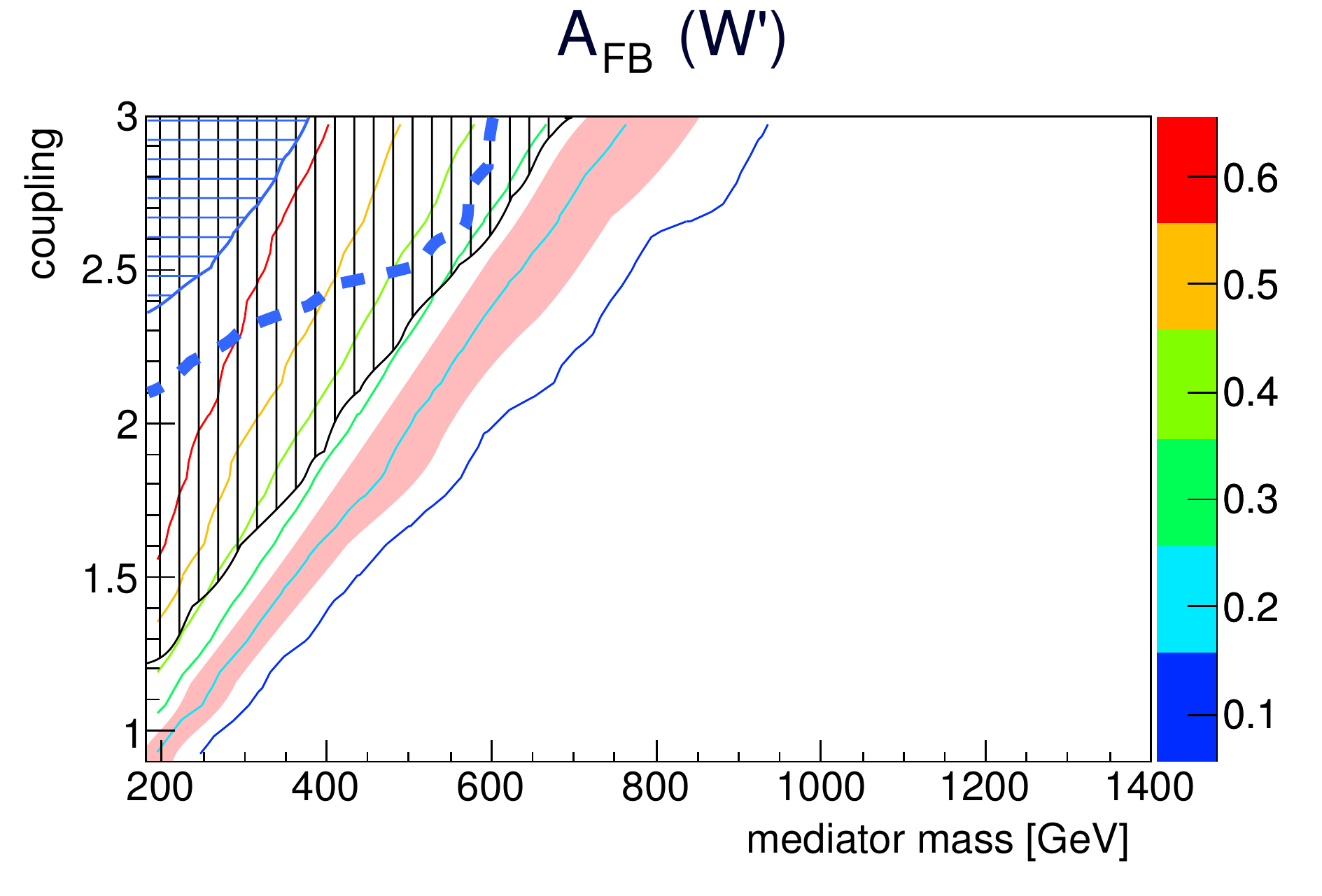}
 \caption{$\afb$ for the $W'$-model; should be read as Fig. \ref{fig:exz_afb}.}
 \label{fig:exw_afb}
\end{center}\end{figure}
We display the asymmetry $\afb$ in Fig.~\ref{fig:exz_afb}.
The pink region describes a $\pm1\sigma$ band from the experimental results, Eq.~\eqref{eq:afb_combined}.
The cross section constraint and the dijet constraint are shown as black- and blue-shaded region, respectively. Also the expected dijet constraint for $5\invfb$ CMS data is displayed as a blue-dashed line.

Here we would like to emphasize the importance of the dijet constraint.
Even in our analysis based on published $36\invpb$ measurements, the dijet constraint slightly exceeds the cross section constraint in a region.
When the CMS accumulates much more data, a vast region in high-coupling range is expected to be judged from the dijet angular distribution measurement.

\subsubsection{$W'$ model}
For $W'$ model, the inclusive asymmetry is displayed in Fig.~\ref{fig:exw_afb}.
A wide region of the experimental band still remains valid in this model.
The dijet constraint is not effective at all because relevant quarks are not $u$-quark but $d$-quark.
\footnote{
The scalar mediator can also give a positive contribution to the top FB asymmetry.
We did the dijet analysis for the models provided in the Refs.~\cite{Shu2010etq,Arhrib2010fba}.
The triplet scalar mediator predicts a positive contribution, while it was found that the dijet constraint is weaker than the result in the $W'$ models. 
}

\section{Conclusion and Discussion}

The Tevatron experiments have reported anomalies in the measurements of the top FB asymmetries. 
This may indicate BSM contributions to the top sector. 
In this letter, the dijet process at the LHC was studied in several BSM models.
The anomalous FB asymmetry is explained by the flavor-changing couplings in the $t$-mediator models. 
It is emphasized that the interactions inevitably contribute to the dijet production cross section at the one-loop level. 
The BSM contributions become prominent in large dijet invariant mass and small rapidity regions. 
It was found that they are comparable to the SM contribution when the coupling $\lambda$ is as large as $O(1)$.
The present LHC results have already constrained the region $\lambda \gsim 2.5$--$3$ for the $Z'$ models, while there is no bound for the other BSM models. 

Although the constraints look less restrictive, our used data correspond only to the integrated luminosity $36\invpb$, and the situation will improve significantly, because the luminosity of the LHC increases rapidly. Suppose that the inclusive FB asymmetry is larger than 0.2, the deviation from the SM will be measured in the dijet cross section if $Z'$ has a mass larger than 800 GeV for $5\invfb$ with crude assumptions. 

The discovery potential of the BSM models through the dijet signals is expected to be improved as the luminosity and the center-of-mass energy are upgraded in the LHC. 
Since the experimental uncertainties crucially determine the sensitivity, it is difficult to predict something definite unless the real data will be published. Nevertheless, if the $t$-mediator models contribute to the top FB asymmetry, they could be accessed at the LHC through the dijet measurements especially when the mediator is heavy. We retain detailed studies for future works.

Let us comment on the acceptance effects. 
The acceptance of the $\ttbar$ reconstruction is assumed to be the same as the SM in this letter. When the data is unfolded by the acceptance, the CDF and D0 results assume that the top quarks have the same event distributions especially for the rapidity as the SM. However, the acceptance can decrease in BSM models \cite{Gresham2011mon,Jung2011tqa,Grinstein2011FSS}.
In the vector $t$-mediator models, the top tends to lie in a large rapidity region, where the acceptance of the semi-leptonic top decays drops rapidly. This is prominent if the vector mediator is relatively light. Thus, the top FB asymmetry is considered to be diluted in small mass regions.
This can improve the sensitivity to the BSM models in the dijet signal for the regions, because the coupling constant is required to be larger to explain $\afb$.

Lastly, we mention the other LHC signatures of the BSM models. 
The measurements of the charge asymmetry of $t\bar t$ at the LHC \cite{CMSPASTOP11014,ATLAS2011106} are tightly correlated with the top FB asymmetry at the Tevatron\cite{Kuhn1998cai,Kuhn1999cah,Antunano2008tqa,Xiao2011eca,Hewett2011aml,Aguilar-Saavedra2011ptt,Shu2010etq,Aguilar-Saavedra2011smt}. Although the current results from the LHC are still dominated by experimental uncertainties, the measurements will provide an important test for the Tevatron results. Since the sensitivity to the BSM effects of the dijet process depends on the models, it may be possible to distinguish them by measuring the dijet production cross section if the anomaly will be confirmed.

The BSM models also contribute to the top measurements at the LHC such as those of the $\ttbar$ production cross section and the same-sign top production. They provide independent signatures of the BSM models for the top FB asymmetry. Since the processes are different, the sensitivities depend on details of the models. In fact, although the simple $Z'$ models for the top FB asymmetry are already excluded by the measurement of the same-sign top production at the CMS \cite{Chatrchyan2011sst}, the constraint could be evaded, depending on the flavor structure of the $Z'$ couplings (see e.g.~\cite{Grinstein2011FSS}). Thus, it is expected that the dijet analysis in this letter will provide a complimentary test of the BSM models in the LHC. 

\section*{Acknowledgment}
This work was supported by Grand-in-Aid for Scientific research from the Ministry of Education, Science, Sports, and Culture (MEXT), Japan, No. 23740172 (M.E.), and by JSPS Grant-in-Aid for JSPS Fellows (S.I.).
This work was also supported by World Premier International Center Initiative (WPI Program), MEXT, Japan.

%

\dummybibliography{ExpResLHC,ExpResTevatron,TopAFB,HEPComputing,FlavorSym}
{\small
\bibliography{afb_rpv_combined}
}
\end{document}